\documentclass[12pt,letterpaper]{article}
\usepackage{graphics}
\usepackage{amssymb,epsfig,amsmath,euscript,array}
\usepackage{axodraw}
\usepackage{subfigure}
\usepackage{color}
\usepackage{amsfonts}
\usepackage[x11names]{xcolor}
\usepackage{epsfig}
\usepackage{graphicx}
\usepackage{nicefrac}
\usepackage{hhline}

\newcommand{\Eqref}[1]{Eq.~\eqref{#1}}

\makeatletter
\@addtoreset{equation}{section}
\makeatother

\newcounter{multieqs}

\newcommand{\fssd}[1]{#1\!\!\!\!/}

\def\bd{\begin{document}}
\def\ed{\end{document}}
\def\nn{\nonumber}
\def\bea{\begin{eqnarray}}
\def\eea{\end{eqnarray}}
\let\bm=\bibitem
\let\la=\label

\begin{document}

\hfill{DESY 07-207}\\[-0.9cm]

\hfill{OUTP-0715P}\\[-0.9cm]

\hfill{IPPP/07/93}\\[-0.9cm]

\hfill{DCPT/07/186}\\[-0.9cm]

\vspace{20pt}

\begin{center}

{\Large \bf Laser experiments explore the hidden sector} \\[1.5ex]

\vspace{30pt}

{M.~Ahlers$^1$, H.~Gies$^2$, J.~Jaeckel$^{2,3}$, J.~Redondo$^4$ and A.~Ringwald$^4$}

{\small \em
{}$^1$Rudolf Peierls Centre for Theoretical Physics, University of Oxford,\\
1 Keble Road, Oxford OX1 3NP, United Kingdom \\[1ex]
{}$^2$Institute for Theoretical Physics, Heidelberg University,\\
Philosophenweg 16, D-69120 Heidelberg, Germany\\[1ex]
{}$^3$Institute for Particle Physics Phenomenology, Durham University,\\ Durham DH1 3LE, United Kingdom\\[1ex]
{}$^4$Deutsches Elektronen-Synchrotron DESY,\\
Notkestrasse 85, D-22607  Hamburg, Germany}

\vspace{40pt}

{\bf Abstract}
\end{center}

\noindent
Recently, the laser experiments BMV and GammeV, searching for light shining through walls,
have published data and
calculated new limits on the allowed masses and couplings
for axion-like particles. In this note we point out that these experiments can serve to constrain a
much wider variety of hidden-sector particles such as,
{\it e.g.}, minicharged particles and hidden-sector photons. The new experiments improve the existing
bounds from the older BFRT experiment by a factor of two.
Moreover, we use the new PVLAS constraints on a possible rotation and ellipticity of light after it
has passed through a strong magnetic field to constrain
pure minicharged particle models. For masses $\lesssim 0.05 \,{\rm{eV}}$, the charge is now restricted
to be less than $(3-4)\times 10^{-7}$ times the electron
electric charge. This is the best laboratory bound and comparable to bounds
inferred from the energy spectrum of the cosmic microwave background.

\setcounter{page}{0}
\thispagestyle{empty}
\newpage
\section{Introduction}
Most extensions of the Standard Model, notably the ones based on
string theory, predict a so-called ``hidden sector'' of particles
which transform trivially under the Standard model gauge group and therefore do
 not directly interact via the Standard Model forces with the
known particles from the ``visible sector''.
Those hidden-sector particles typically interact via feeble,
gravity-like interactions with the Standard Model. For example in
supersymmetric extensions of the Standard Model, supersymmetry (SUSY)
breaking is often assumed to be generated at a high scale by
hidden-sector dynamics and then communicated to the minimal
supersymmetric Standard Model (MSSM) sector by such forces.

Frequently, it is assumed that all the particles in the hidden sector
are very heavy.  This is, however, not necessarily the case. For
example, the hidden sector gauge group may contain unbroken U(1) gauge
factors, corresponding to additional massless spin-one bosons,
potentially mixing with the photon. Moreover, chiral symmetries may be
responsible for keeping some matter particles from the hidden sector
very light. Finally, if these particles are charged under the above
U(1), they {typically} acquire a small electric charge due to the above
mentioned mixing phenomenon. Thus, light minicharged particles (MCPs)
arise naturally within hidden sectors.

Laser experiments provide a powerful laboratory tool to shed light on
hidden sectors with potentially tiny couplings to photons. Laser
polarization experiments --- such as BFRT~\cite{Cameron:1993mr},
PVLAS~\cite{Zavattini:2005tm}, and Q\&A~\cite{Chen:2006cd}, where
linearly polarized laser light is sent through a transverse magnetic
field, and changes in the polarization state are searched for --- are
sensitive to axion-like particles
(ALPs) \cite{Maiani:1986md,Raffelt:1987im}, to MCPs~\cite{Gies:2006ca}, to hidden sector U(1)
bosons~\cite{Ahlers:2007rd,Antoniadis:2007sp} and to other very weakly
interacting sub-eV particles (WISPs) such as
chameleons~\cite{Brax:2007hi} and the like \cite{Jaeckel:2006xm}.
Light-shining-through-walls (LSW) experiments, such as
BFRT~\cite{Cameron:1993mr,Ruoso:1992nx}, are another powerful tool to
search for WISPs. Here, laser light is shone onto a wall, and one
searches for photons that re-appear behind the wall. Vacuum
oscillations of photons into hidden-sector photons with sub-eV masses
would lead to a non-vanishing regeneration rate~\cite{Okun:1982xi}. In
the presence of a magnetic field, photons can oscillate into
ALPs~\cite{Sikivie:1983ip,Anselm:1986gz,Gasperini:1987da,VanBibber:1987rq}
or into massless hidden-sector photons coupling to light hidden-sector
particles~\cite{Ahlers:2007rd}, which can be reconverted into photons
behind the wall by another magnetic field.

Recently, the laser polarization experiment
PVLAS~\cite{Zavattini:2007ee} and the LSW experiments
BMV~\cite{Robilliard:2007bq} and GammeV~\cite{Chou:2007zz} published
new results. All three experiments found no significant signs of a
signal and put corresponding limits on the coupling of a hypothetical
ALP to a photon. But from the optical data, much more information
about hidden-sector scenarios can be extracted. Further results
can be expected from similar experiments such as ALPS
\cite{Ehret:2007cm}, LIPSS \cite{Afanasev:2006cv}, OSQAR
\cite{OSQAR}, and PVLAS LSW~\cite{PVLASLSW} in the near future. It is the purpose of this note to
explore the parameter space of hidden-sector photons and minicharged
particles in the light of the new data.

\section{Minicharged particles}\label{MCPs}

The first hypothesis which we will confront with the new data involves
light particles with mass $m_\epsilon$ and small charges $\epsilon e$
under the electromagnetic U(1): minicharged particles (MCPs). Within
the context of laser experiments we expect to be sensitive
only to $m_\epsilon$ much smaller than the electron mass where
$\epsilon$ should be kept $ \ll 1$. As an effective low-energy
theory, we will assume that a standard minimal coupling between the
MCPs and the photon exists; moreover, we consider both fermionic Dirac
spinor MCPs as well as complex scalar MCPs.

If this low-energy theory was valid even at solar energy scales
$\sim$keV, the astrophysics of horizontal branch stars would already imply strong
constraints on the MCP parameters, resulting in $\epsilon\leq 2\times
10^{-14}$, for $m_{\epsilon}$ below a few keV~\cite{Davidson:2000hf}.
However, if the MCP parameters are generated
by the hidden sector at scales much below $\mathcal O(\text{keV})$,
solar physics {can be un}affected by these hidden-sector degrees of
freedom, which instead can become visible in laboratory experiments at
the {sub-eV} scale. A particular scenario entailing such a mechanism is
the Masso-Redondo (MR) model~\cite{Masso:2006gc}, involving MCPs as well as hidden-sector
photons, to be discussed in Sect.~\ref{hiddenphotons+MCP}. A model
that can give
rise to MCPs as low-energy degrees of freedom without requiring
hidden-sector photons has been worked out
in \cite{Batell:2005wa} within the context of warped extra
dimensions.

%%%%%%%%%%%%%%%%%%%%%%%%%%%%
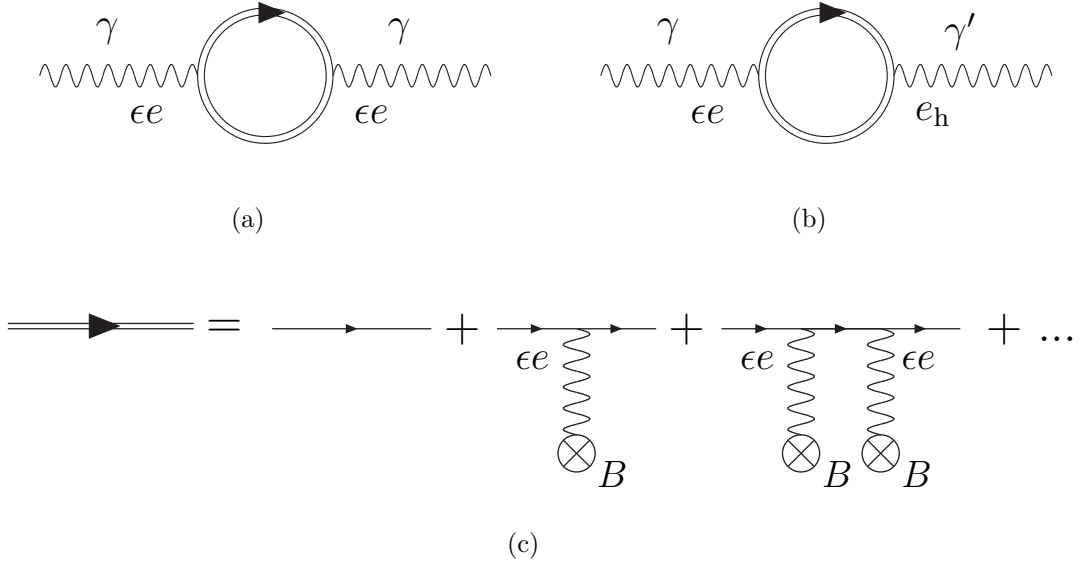
\begin{figure}[t]
\begin{center}
%\hspace{-1cm}
\subfigure[]{
\scalebox{0.85}[0.85]{
\begin{picture}(190,120)(0,40)
\Photon(0,90)(70,90){5}{7.5}
\Text(30,110)[c]{\scalebox{1.5}[1.5]{$\gamma$}}
\Text(160,110)[c]{\scalebox{1.5}[1.5]{$\gamma$}}
\Text(40,73)[l]{\scalebox{1.5}[1.5]{$\epsilon e$}}
\Text(140,73)[l]{\scalebox{1.5}[1.5]{$\epsilon e$}}
\CArc(100,90)(30,0,180)
\CArc(100,90)(30,180,360)
\CArc(100,90)(27,0,180)
\CArc(100,90)(27,180,360)
{\SetWidth{3}
\ArrowLine(100,118)(106,118)}
\SetOffset(0,0)
\Photon(130,90)(200,90){5}{7.5}
\end{picture}
}
\label{convertAA}}
\hspace{1cm}
\subfigure[]{
\scalebox{0.85}[0.85]{
\begin{picture}(190,120)(0,40)
\Photon(0,90)(70,90){5}{7.5}
\Text(30,110)[c]{\scalebox{1.5}[1.5]{$\gamma$}}
\Text(160,110)[c]{\scalebox{1.5}[1.5]{$\gamma'$}}
\Text(40,73)[l]{\scalebox{1.5}[1.5]{$\epsilon e$}}
\Text(140,73)[l]{\scalebox{1.5}[1.5]{$e_\mathrm{h}$}}
\CArc(100,90)(30,0,180)
\CArc(100,90)(30,180,360)
\CArc(100,90)(27,0,180)
\CArc(100,90)(27,180,360)
{\SetWidth{3}
\ArrowLine(100,118)(106,118)}
\SetOffset(0,0)
\Photon(130,90)(200,90){5}{7.5}
\end{picture}
}
\label{convertBB}}
\subfigure[]{\begin{picture}(190,100)(0,20)
\SetScale{1}
\SetOffset(-100,90)
\Line(0,0)(70,00)
\Line(0,2)(70,2)
{\SetWidth{3} \ArrowLine(34,1)(39,1)}
\Text(75,0)[l]{\scalebox{1.5}[1.5]{$=$}}
\SetOffset(00,90)
 \ArrowLine(0,0)(60,0)  \Text(65,0)[l]{\scalebox{1.5}[1.5]{$+$}}
\SetOffset(85,90)
 \ArrowLine(0,0)(30,0) \ArrowLine(30,0)(60,0) \Text(65,0)[l]{\scalebox{1.5}[1.5]{$+$}}
 \Photon(30,0)(30,-40){5}{5} \CArc(30,-47)(7,0,360) \Line(35,-52)(25,-42) \Line(35,-42)(25,-52)
\Text(38,-55)[l]{\scalebox{1.2}[1.2]{$B$}}
\SetOffset(170,90)
 \ArrowLine(0,0)(30,0) \ArrowLine(30,0)(60,0)  \Photon(30,0)(30,-40){5}{5} \CArc(30,-47)(7,0,360)
\Line(35,-52)(25,-42)
 \Line(35,-42)(25,-52)  \Text(38,-55)[l]{\scalebox{1.2}[1.2]{$B$}}
\SetOffset(200,90)
\ArrowLine(0,0)(30,0) \ArrowLine(30,0)(60,0)  \Photon(30,0)(30,-40){5}{5} \CArc(30,-47)(7,0,360)
\Line(35,-52)(25,-42) \Line(35,-42)(25,-52)
\Text(38,-55)[l]{\scalebox{1.2}[1.2]{$B$}}
 \Text(70,0)[l]{\scalebox{1.5}[1.5]{$+\ ...$}}
\SetOffset(0,0)
\Text(92,78)[l]{\scalebox{1.275}[1.275]{{$\epsilon e$}}}
\Text(177,78)[l]{\scalebox{1.275}[1.275]{{$\epsilon e$}}}
\Text(238,78)[l]{\scalebox{1.275}[1.275]{{$\epsilon e$}}}
\end{picture}
\label{convertCC}}
\vspace{.1cm}
\end{center}
\vspace{-0.5cm} \caption[]{\small {
    {The contribution of minicharged particles to the
      polarization tensor \ref{convertAA}. The real part leads to
      birefringence, whereas the imaginary part reflects the absorption
      of photons caused by the production of particle-antiparticle
      pairs.}  The analogous diagram \ref{convertBB} shows how
    minicharged particles mediate transitions between photons and
    hidden-sector photons $\gamma^{\prime}$. Note that the latter diagram is enhanced with respect
    to the first one by a factor {$\sim e_\mathrm{h}/(\epsilon e){=1/\chi}$}.
    The double line represents the complete propagator of the
    minicharged particle in an external magnetic field $B$
    as displayed in \ref{convertCC} \cite{Schwinger:1951nm}. }}
\label{convert-NP}
\end{figure}
%%%%%%%%%%%%%%%%%%%%%%%%%%%%%%%

Within the low-energy effective theory, vacuum fluctuations of MCPs
induce nonlinear and nonlocal self-interactions of the electromagnetic
field (cf.~Fig.~\ref{convertAA}). In polarization experiments where laser photons with a small
amplitude propagate in a strong magnetic field, the equation of motion
for the laser amplitude $a_\mu$ with momentum $k_\mu$,
\begin{equation}
(k^2 g_{\mu\nu} -k_\mu k_\nu + \Pi_{\mu\nu}(k|B)) a^\nu(k)=0,
\label{eomphot}
\end{equation}
involves the polarization tensor $\Pi_{\mu\nu}$ in a magnetic field
$B$~\cite{Toll:1952rq,Tsai:1974fa,Daugherty:1984tr}. The two transverse photon eigenmodes
correspond to polarizations parallel ($\|$ mode) and perpendicular
($\bot$ mode) to the plane spanned by the magnetic field and the
propagation direction. Their dispersion relation in the form of the
eigenvalues gives rise to different vacuum magnetic refractive indices
$n_{\|,\bot}$ and absorption coefficients $\kappa_{\|,\bot}$ for the
two modes, being related to the real and imaginary parts of the
polarization tensor. The magnetized quantum vacuum is birefringent, as
parameterized by $\Delta n=n_\| - n_\bot$, and exhibits dichroism, as
characterized by $\Delta \kappa=\kappa_\| -\kappa_\bot$. These effects
are, of course, present already in the Standard Model, predominantly
owing to electron-positron fluctuations, but MCP contributions can
exceed the Standard-Model effects if the MCP mass $m_\epsilon$ is
sufficiently small~\cite{Gies:2006ca}. The MCP-induced quantities are
\begin{equation}
\Delta n=
-\frac{\epsilon^{2}\alpha}{4\pi}\left(\frac{\epsilon\,eB}{m^{2}_{\epsilon}}\right)^{2}
\Delta I(\lambda), \quad
\Delta \kappa=\frac{1}{2}\epsilon^3 e \alpha \frac{B  }{m_\epsilon}\,
\Delta T(\lambda ),
\label{eq:Delta}
\end{equation}
where $\Delta I=I_{||}-I_\perp$ (and analogously for $\Delta T$)
and, for instance, for a Dirac spinor MCP, we have~\cite{Tsai:1974fa}
\begin{eqnarray}
I_{||,\perp}^{\rm Dsp}(\lambda)\!\!&=&\!\!{2^{\frac{1}{3}}}
\left(\frac{3}{\lambda}\right)^{\frac{4}{3}}
\int^{1}_{0} {\rm d}v\,
\frac{\left[ \left(1-\frac{v^2}{3}\right)_{||}, 
\left(\frac{1}{2}+\frac{v^2}{6}\right)_\perp \right]}{(1-v^{2})^{\frac{1}{3}}}
\tilde{e}^{\prime}_{0}\left[\begin{scriptstyle}-
\left(\frac{6}{\lambda}\frac{1}{1-v^2}\right)^{\frac{2}{3}}\end{scriptstyle}\right],\\
\label{eq:I}
T_{||,\perp}^{\rm Dsp} &=&
\frac{4\sqrt{3}}{\pi\lambda}
\int\limits_0^1 {\rm d}v\
\frac{\left[ \left(1-\frac{v^2}{3}\right)_{||}, 
\left(\frac{1}{2}+\frac{v^2}{6}\right)_\perp \right]}{(1-v^{2})}
K_{2/3}\left( \frac{4}{\lambda}\frac{1}{1-v^2}\right),  \label{eq:T}
\end{eqnarray} 
where $K_\nu(x)$ is the MacDonald function, and $\tilde{e}'_0(x)$
denotes the derivative of the generalized Airy function
$\tilde{e}_0(x)= \int_0^\infty {\rm d}u \sin( x u - u^3/3)$. For the
corresponding quantities for scalar MCPs, see the appendix of
\cite{Ahlers:2006iz}. The quantity $\lambda$ abbreviates the
dimensionless combination
\begin{equation}
\lambda \equiv  \frac{3}{2} \frac{\omega}{m_\epsilon} \frac{\epsilon e B}{m_\epsilon^2},
\end{equation}
where $\omega$ is the laser frequency.  The above results hold for
magnetic fields which are slowly varying over the scale of the MCP
Compton wavelength $1/m_{\epsilon}$. The formula for the absorption
coefficient difference $\Delta \kappa$ requires the laser frequency to
be above the MCP pair threshold $\omega\geq 2m_{\epsilon}$, and both
formulas assume that a high number of Landau levels can be occupied \cite{Daugherty:1984tr}.
Given that the laser experiments we consider have $\mathcal O(\text{m})$ lengths and
$\mathcal O(\text{eV})$ frequencies we expect our expressions to be valid roughly
for $10^{-7}$~eV$<m_\epsilon<1$~eV.

Laser polarization experiments such as BFRT, PVLAS, and Q\&A search
for vacuum-magnetically-induced $\Delta n$ and $\Delta \kappa$ by
sending a linearly polarized laser beam containing both modes, $\|$
and $\bot$, through a strong magnetic field of $\mathcal O(\text{T})$.
If the magnetized quantum vacuum is birefringent,
$\Delta n\neq0$, the laser light picks up an ellipticity $\psi$; {if it is also
dichroic}, $\Delta \kappa\neq 0$, the amplitudes of the two modes are
depleted differently and the laser polarization undergoes an effective
rotation $\Delta \theta$,
\begin{equation}
\psi=\frac{\omega}{2}\, \ell_B\, \Delta n \, \sin (2\theta), \quad
\Delta\theta = \frac{1}{4}\, \ell_B\, \Delta\kappa \sin (2\theta),
\end{equation}
where $\theta$ is the initial angle of polarization with respect to
the direction of the $B$ field, and $\ell_B$ is the optical path
length inside the magnetic field.
Both expressions have been derived assuming $\psi,\Delta\theta\ll 1$. 
A summary of the experimental parameters can be found in Tabs. \ref{tabrot} and \ref{tabell}.

%%%%%%%%%%%%%%%%%%%%%%%%%%%%%%%%%%%%
\begin{figure}[t]
\centering
\includegraphics[width=0.49\linewidth]{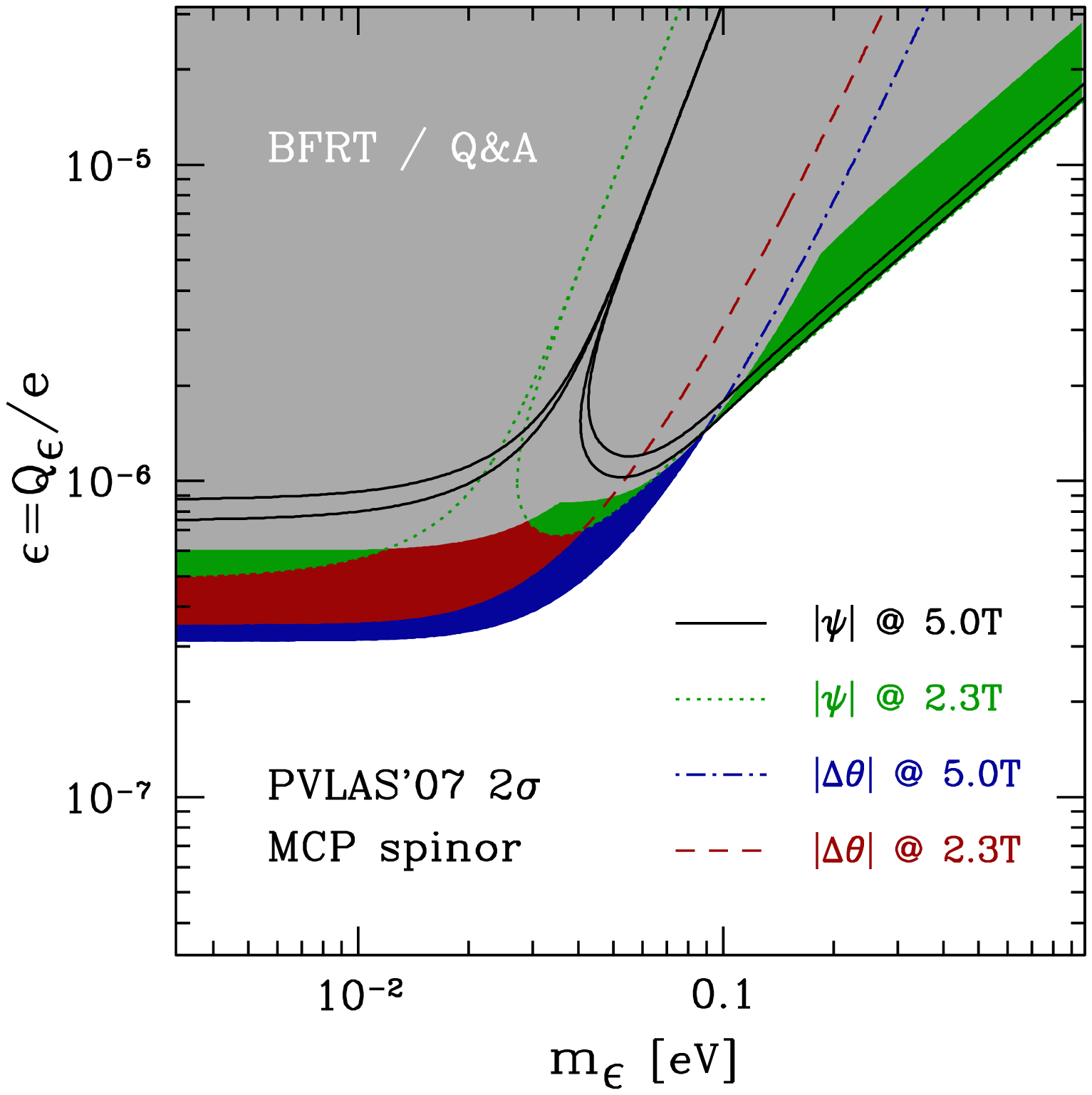}
\hfill
\includegraphics[width=0.49\linewidth]{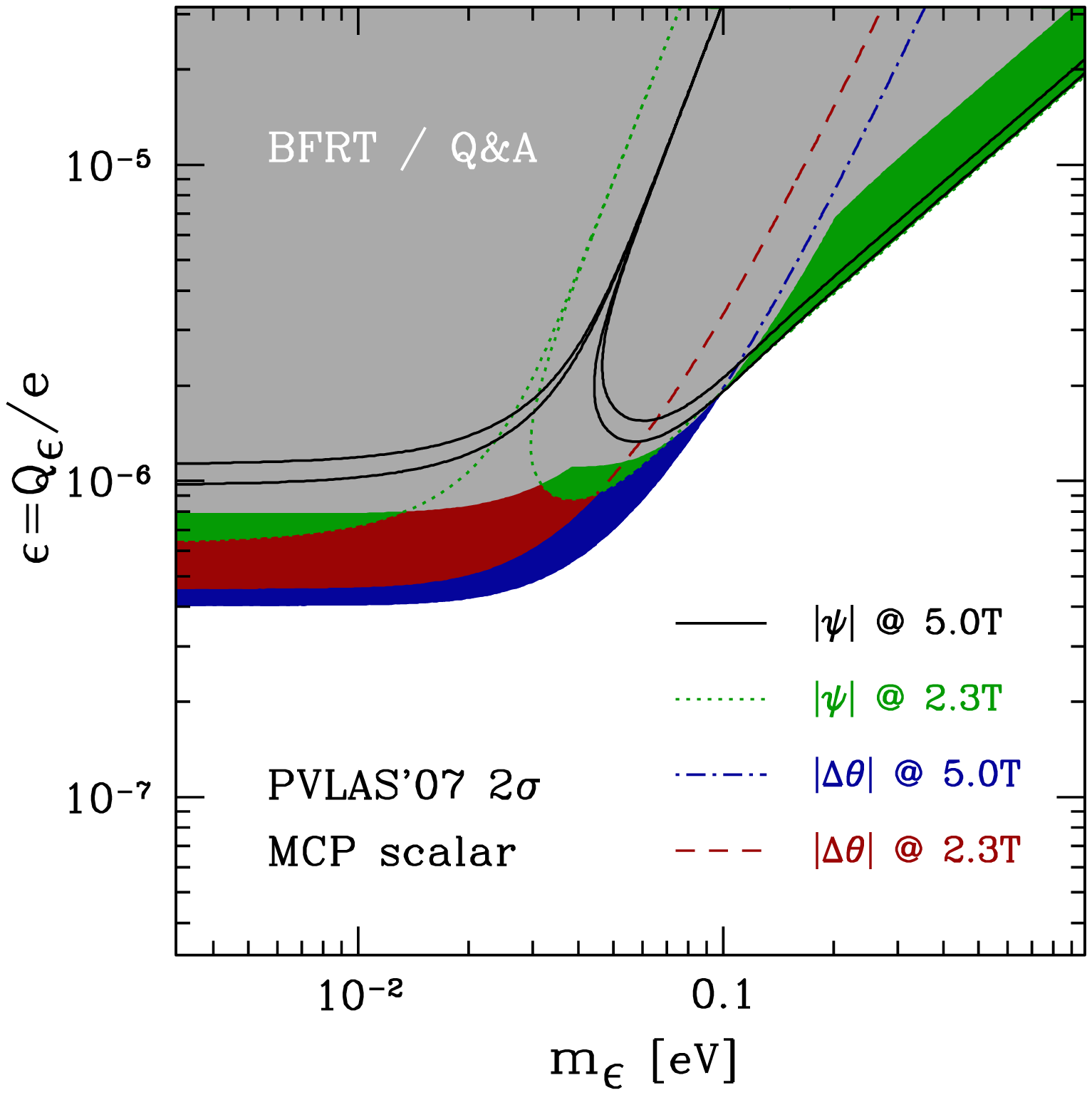}
\caption[]{\small New constraints for the MCP mass $m_\epsilon$ and
  charge fraction $\epsilon$ for Dirac spinor MCPs (left panel) and complex
  scalar MCPs (right panel) as deduced from the new PVLAS data
  \cite{Zavattini:2007ee}.
  Also shown are the constraints obtained from the BFRT~\cite{Cameron:1993mr}
  and Q\&A~\cite{Chen:2006cd} data, as derived in Ref.~\cite{Ahlers:2006iz}.
  The constraints result from the absence of
  ellipticity $\psi$ and rotation $\Delta\theta$ signals, with the
  shaded parameter space being excluded
  at 95\% C.L.. The 2$\sigma$ region for the ellipticity signal $\psi$ at
  $B=5$~T is marked with solid lines; 
  apart from a marginally allowed region at larger masses, this signal is excluded in the MCP
  scenario.}\label{figMCP}
\end{figure}
%%%%%%%%%%%%%%%%%%%%%%%%%%%%%%%%%%%%

The PVLAS experiment{\footnote{In the PVLAS setup, $\theta$ is slowly changing in time covering
all values from $0$ to $2\pi$. For our exclusion bounds we are taking just $2\theta=\pi/2$.}}
has recently published new constraints on $\Delta
n$ and $\Delta \kappa$, resulting from ellipticity $\psi$ and rotation
$\Delta \theta$ measurements at $B=2.3$~T and $B=5$~T. No signals were
observed for $\Delta \theta(B=2.3\ \text{T})$, $\Delta
\theta(B=5\ \text{T})$, and $\psi(B=2.3\ \text{T})$; a signal was
present in a $\psi(B=5\ \text{T})$ measurement which is likely to
result from an instrumental artifact.

%%%%%%%%%%%%%%%%%%%%%%%%%%%%%%%%%%%%
\begin{table}[t]\centering
\renewcommand{\arraystretch}{2.0}\scriptsize
\begin{tabular}{lcccccc}\hline
Experiment&$\omega$[eV]&$N_\text{pass}$&$B_{\rm{range}}$[T]&$\ell_B$[m]&$|\Delta \theta|$~[nrad]\\
\hline
BFRT&$2.47$&{$254$}&{$2.6-3.9$}&{$8.8$}&$<0.60$~\text{(95\% C.L.)}\\
Q\&A&$1.17$&{$18700$}&$0-2.3$&{$1.0$}&$<10$~\text{(95\% C.L.)}\\
PVLAS ``low field''&$2.33$&45000&$0-2.3$&{$1.0$}&$<10$~\text{(95\% C.L.)}\\
PVLAS ``high field''&$2.33$&45000&$0-5.0$&{$1.0$}&$<12$~\text{(95\% C.L.)}\\
\hline
\end{tabular}
\caption[]{\small Parameters of the polarization experiments searching for a possible rotation 
$\Delta \theta$ of the polarization after passage
through a magnetic field. $\omega$ is the frequency of the laser light, $N_\text{pass}$ is the number 
of passes through the cavity of length $\ell_{B}$.
$B_{\rm{range}}$ denotes the range over which the magnetic field projected on a fixed direction 
perpendicular to the laser beam is varied.
In the BFRT experiment this range is
achieved by ramping the magnetic field between two different field strengths while keeping the direction 
of the magnetic field fixed. In the Q\&A and PVLAS experiments the
magnetic field is rotated while keeping the overall field strength constant.
}\label{tabrot}
\end{table}
%%%%%%%%%%%%%%%%%%%%%%%%%%%%%%%%%%%%%
%%%%%%%%%%%%%%%%%%%%%%%%%%%%%%%%%%%%
\begin{table}[t]\centering
\renewcommand{\arraystretch}{2.0}\scriptsize
\begin{tabular}{lcccccc}\hline
Experiment&$\omega$[eV]&$N_\text{pass}$&$B_{\rm{range}}$[T]&$\ell_B$[m]&$|\psi|$~[nrad]\\
\hline
BFRT&$2.47$&$34$&{$2.6-3.9$}&{$8.8$}&$<2.0$~\text{(95\% C.L.)}\\
PVLAS ``low field''&$2.33$&45000&$0-2.3$&{$1.0$}&$<14$~\text{(95\% C.L.)}\\
PVLAS ``high field''&$2.33$&45000&$0-5.0$&{$1.0$}&$90\pm9$\\
\hline
\end{tabular}
\caption[]{\small As in Tab. \ref{tabrot}, but searches for a possible ellipticity of the polarization.
}\label{tabell}
\end{table}
%%%%%%%%%%%%%%%%%%%%%%%%%%%%%%%%%%%%%

These measurements translate into new bounds for the MCP parameters
shown in Fig.~\ref{figMCP} for Dirac spinors (left panel) and
complex scalars (right panel). The shaded parameter regions
show  the excluded domain at 95\%C.L.. For instance for Dirac
spinor MCPs, we find that $\epsilon\lesssim 3\times 10^{-7}$ for
$m_\epsilon < 30$~meV. This bound is indeed of a similar size as a
cosmological MCP bound which has recently been derived from a
conservative estimate of the distortion of the energy spectrum of
the cosmic microwave background \cite{Melchiorri:2007sq}. Hence,
laboratory experiments begin to enter the hidden-sector parameter
regime which has previously been accessible only to cosmological and
astrophysical considerations.

Incidentally, the anomalous ellipticity signal $\psi(B=5\ \text{T})$
(solid lines) is only marginal\-ly compatible with the other data
for larger masses $m_{\epsilon}\gtrsim 0.1$~eV within the MCP
hypothesis.  This is in line with the PVLAS
interpretation that this signal results from an instrumental
artifact.

\section{Massive hidden-sector photons}\label{hiddenphotons}

The next hypothesis which we will confront with data is based on the
assumption that the low-energy dynamics involves, in addition to the
familiar massless electromagnetic {U(1)$_{_\mathrm{QED}}$}, another
hidden-sector {U(1)$_\mathrm{h}$} under which all Standard Model
particles have zero charge.  The most general renormalizable
Lagrangian describing these two U(1) gauge groups at low energies is
\begin{equation}
\label{lagrangian}
{\mathcal{L}}= -\frac{1}{4} F^{\mu\nu}F_{\mu\nu}-\frac{1}{4}B^{\mu\nu}B_{\mu\nu}
-\frac{1}{2}\chi\,F^{\mu\nu}B_{\mu\nu}  +\frac{1}{2}m_{\gamma^\prime}^2 B_\mu B^\mu,
\end{equation}
where $F_{\mu\nu}$ is the field strength tensor for the ordinary
electromagnetic {U(1)$_{_\mathrm{QED}}$} gauge field $A^{\mu}$, and
$B^{\mu\nu}$ is the field strength for the hidden-sector
{U(1)$_\mathrm{h}$} field $B^{\mu}$.  The first two terms are the
standard kinetic terms for the photon and hidden-sector photon fields,
respectively. Because the field strength itself is gauge invariant for
U(1) gauge fields, the third term is also allowed by gauge and Lorentz
symmetry.  This term corresponds to a non-diagonal kinetic term, a
so-called kinetic mixing~\cite{Holdom:1985ag,Foot:1991kb}. From the
viewpoint of a low-energy effective Lagrangian, $\chi$ is a completely
arbitrary parameter. Embedding the model into a more fundamental
theory, it is plausible that $\chi=0$ holds at a high-energy scale
related to the fundamental theory. However, integrating out the heavy
quantum fluctuations generally tends to generate non-vanishing~$\chi$
at low scales. Indeed, kinetic mixing arises quite generally both in
field theoretic~\cite{Holdom:1985ag,Foot:2007cq} as well as in string
theoretic~\cite{Dienes:1996zr,Lust:2003ky,Abel:2003ue,Blumenhagen:2006ux,Abel:2006qt,Kim:2007wj}
setups, and typical predicted values
for $\chi$ range between $10^{-16}$ and $10^{-4}$.
The last term in the Lagrangian~(\ref{lagrangian}) accounts for a
possible mass of the paraphoton. This may arise via a Higgs or, alternatively,
via a Stueckelberg~\cite{Stueckelberg:1938}
mechanism.

%%%%%%%%%%%%%%%%%%%%%%%%%%%%%%%%%%%%
\begin{table}[t]\centering
\renewcommand{\arraystretch}{2.0}\scriptsize
\begin{tabular}{lcccccccccc}\hline
Experiment&$\omega$[eV]&$N_\text{pass}$&$B$[T]&$\ell_1$[m]&$\ell_2$[m]&$L_1$[m]&$L_2$[m]&$N_0$&$\eta$&$N_{95\%}$\\
\hline
BFRT&$2.47$&200&$3.7$&$4.4$&$4.4$&$11$&$6.5$&$7.8\times10^{18}$~Hz&$0.055$&$0.018$~Hz\\
BMV&$1.17$&-&$12.3$&$0.365$&$0.365$&$20.0$&$1.0$&$6.7\times10^{22}$&$0.5$&$3.09$\\
GammeV ``centre''&$2.33$&-&$5.0$&$3.1$&$2.9$&$5.4$&$7.2$&$6.6\times10^{23}$&$0.33$&$3.69$\\
GammeV ``edge''&$2.33$&-&$5.0$&$5.0$&$1.0$&$7.3$&$7.1$&$6.4\times10^{23}$&$0.33$&$2.05$\\
\hline
\end{tabular}
\caption[]{\small Parameters of LSW experiments. Here $\omega$ is the frequency of the 
laser beam, $N_{\rm{pass}}$ the number of passes through the cavity (if present)
and $B$ the magnetic field strength. $\ell_{1}$ and $\ell_{2}$ are the lengths of the magnetized 
regions of the production and regeneration sides of the experiments
whereas $L_{1}$ and $L_{2}$ are the total lengths on both sides including regions without magnetic 
field. $N_0$ is the photon number or rate of the laser beam, $\eta$ the
quantum efficiency of the detector and $N_{95\%}$ the upper limit on the detected number of photons.
We follow Ref.~\cite{Feldman:1997qc} in deriving
the 95\% C.L.~upper limits ($N_{95\%}$) for the BMV and GammeV results.
}\label{tab1}
\end{table}
%%%%%%%%%%%%%%%%%%%%%%%%%%%%%%%%%%%%%

Let us now switch to a field basis in which the prediction of photon $\leftrightarrow$
hidden-sector photon oscillations becomes apparent.  In fact, the
kinetic terms in the Lagrangian~(\ref{lagrangian}) can be diagonalized
by a shift
\begin{equation}
\label{shift}
B^{\mu}\rightarrow \tilde{B}^{\mu}-\chi A^{\mu}.
\end{equation}
Apart from a multiplicative renormalization of the electromagnetic
gauge coupling, $e^2\rightarrow e^2/(1-\chi^2)$, the visible-sector
fields remain unaffected by this shift and one obtains a non-diagonal
mass term that mixes photons with hidden-sector photons,
\begin{equation}
\label{masssimple}
{\mathcal{L}}=-\frac{1}{4} F^{\mu\nu}F_{\mu\nu}-\frac{1}{4}\tilde{B}^{\mu\nu}\tilde{B}_{\mu\nu}
+\frac{1}{2}m_{\gamma^\prime}^2 \left(\tilde{B}^{\mu}\tilde{B}_{\mu}-2\chi \tilde{B}^{\mu}A_{\mu}+
\chi^2 A^{\mu}A_{\mu}\right) ,
\end{equation}
where we have absorbed an irrelevant factor of $\sqrt{1-\chi^2}$ in both $A_\mu$ and $1/\chi$.  
Therefore, in analogy to neutrino flavour
oscillations, photons may oscillate in vacuum into hidden-sector
photons.  These oscillations and the fact that hidden-sector photons
do not interact with ordinary matter forms the basis of the
possibility to search for signals of hidden-sector photons in LSW
experiments. 
The probability for a photon to pass through the wall and to arrive at
the detector is~\cite{Okun:1982xi,Ahlers:2007rd},
\begin{equation}
\label{onedimensional}
P_\text{trans} =
16 \chi^4
{\left[\sin\left(\frac{\Delta k L_{1}}{2}\right)
\sin\left(\frac{\Delta k L_{2}}{2}\right)\right]^2},
\end{equation}
where $L_{1}$ ($L_{2}$) is the distance between the laser and the wall{\footnote{More precisely, it is the
length between the last optical device that redirects the laser beam light into the wall and the wall
itself. See Appendix B of Ref.~\cite{Ahlers:2007rd}.}}
(the wall and the detector), and
\begin{equation}
\Delta k=\omega-\sqrt{\omega^{2}-m_{\gamma^\prime}^2}\approx m_{\gamma^\prime}^2/(2\omega)
,\hspace{6ex} {\rm for\ }\ m_{\gamma^\prime}\ll \omega ,
\end{equation}
is the momentum difference between the photon and the hidden-sector
photon, expressed in terms of the energy of the laser photons,
$\omega$. If a cavity is used on the production side, {\it i.e.},
before the wall, \Eqref{onedimensional} receives an additional factor
of $(N_\text{pass}+1)/2$, where $N_\text{pass}\gg 1$ is the number of
passes the light makes through the cavity; also, the length $L_1$ then
has to be replaced by the path length $\ell_1$ inside the cavity. The
LSW transition probability in \Eqref{onedimensional} is actually
independent of the magnetic field, since the mixing arises from the
mass term. The expected rate of observed photons {in addition
involves} the total initial photon rate $N_0$ and the detection
efficiency $\eta<1$, $N = \eta N_0 P_\text{trans}$. The experimental
parameters of BFRT~\cite{Cameron:1993mr}, BMV~\cite{Robilliard:2007bq}
and GammeV~\cite{Chou:2007zz}, as relevant for the search to
paraphotons, are summarized in Table~\ref{tab1}.

\begin{figure}[t]
\centering
\includegraphics[width=0.49\linewidth]{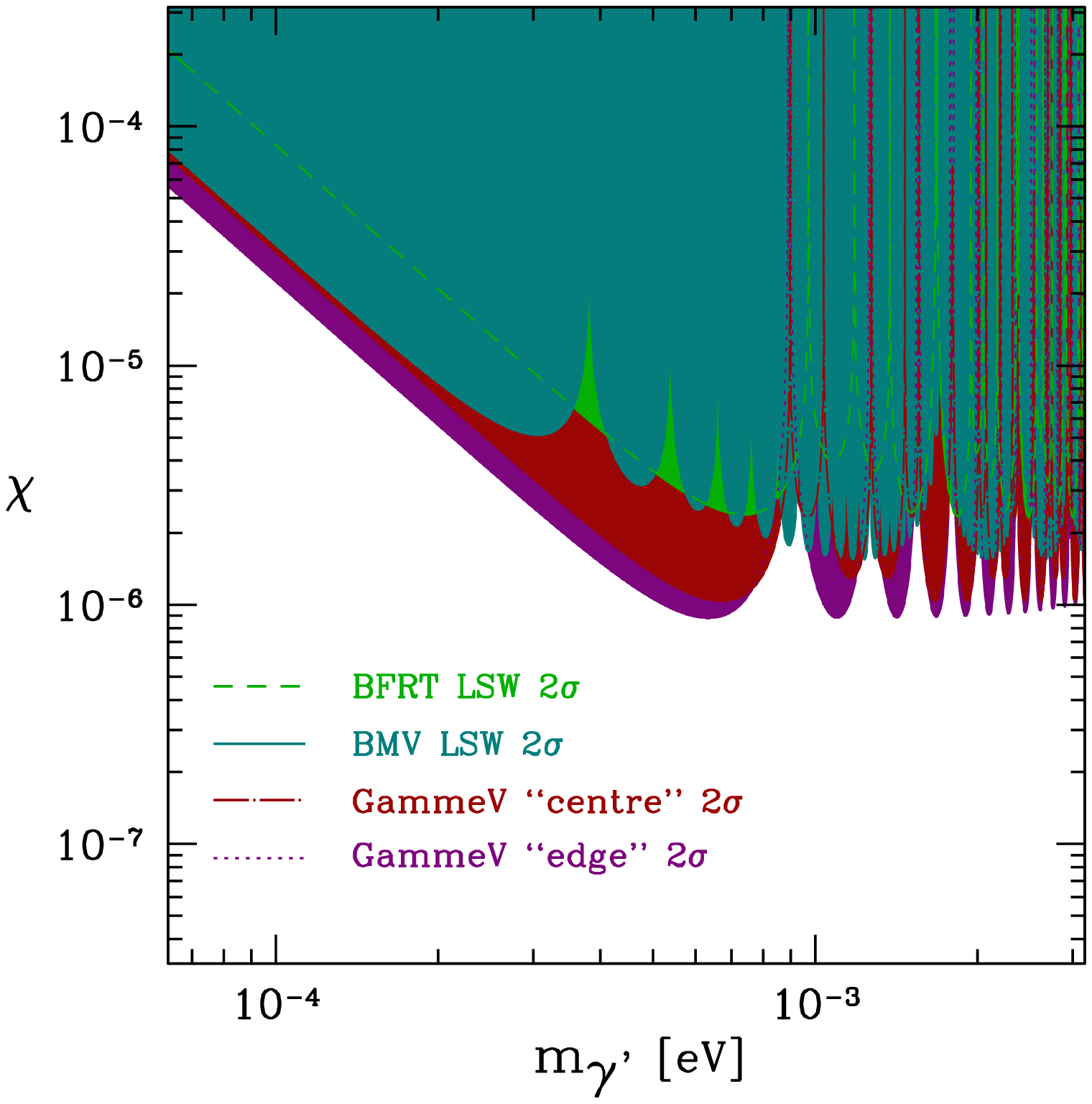}
\hfill
\includegraphics[width=0.49\linewidth]{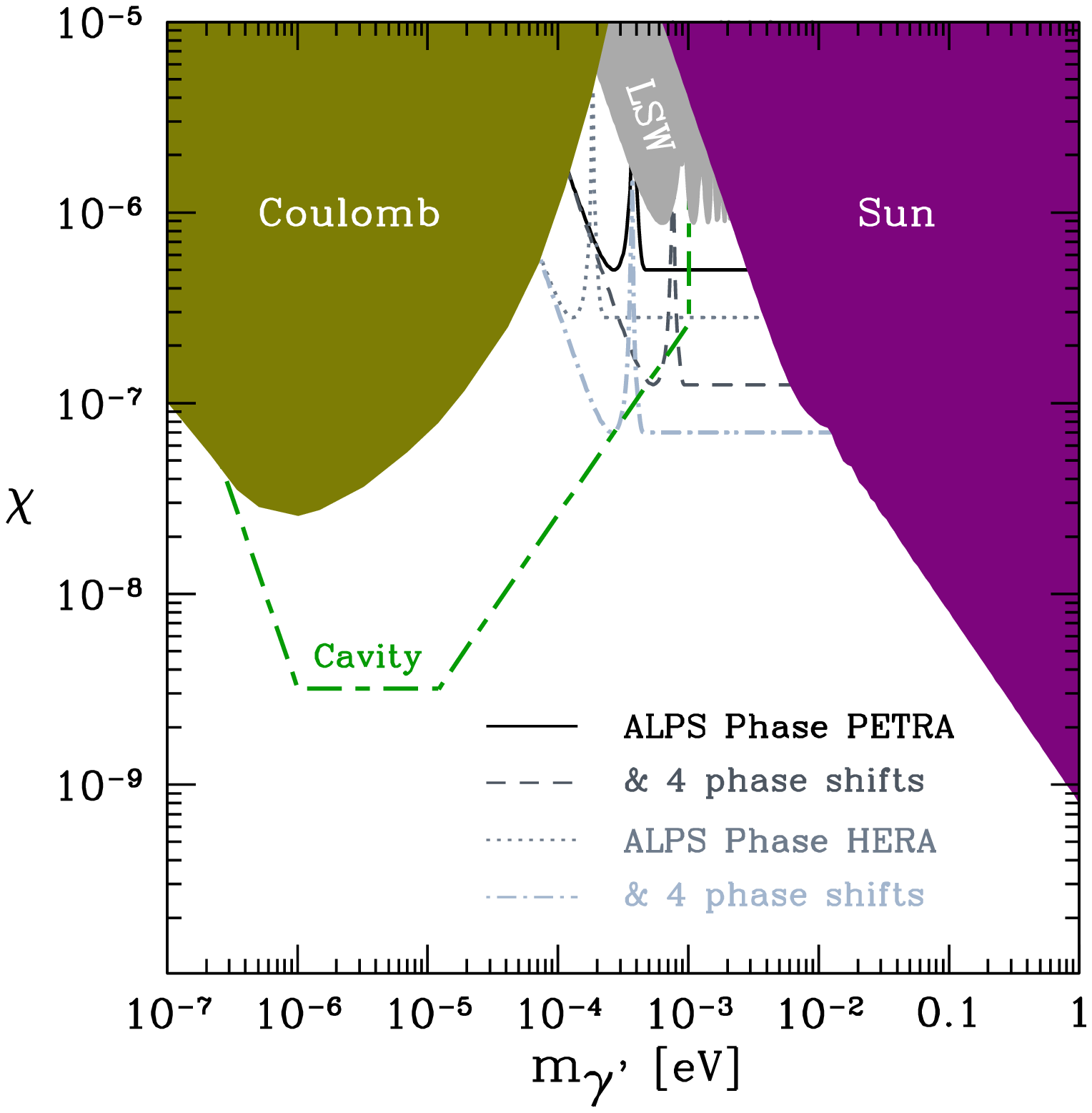}
\caption[]{\small Limits on hidden-sector photons $\gamma^\prime$ mixing
with the photon from searches for photon regeneration in LSW
experiments. {\bf Left panel:} New limits from the non-observation in the LSW experiments BMV and
GammeV compared to the old BFRT results. The bounds relax by a factor $1\sim 3$ in the MR
model \cite{Masso:2006gc} depending on $m_{\epsilon}$. See
Sect.~\ref{hiddenphotons+MCP} for details.
{\bf Right panel:} Forecast of future experiments searching for $\gamma^\prime$s.
The current limit from LSW experiments (grey shaded) can be extended by dedicated
experiments exploiting high power, $\sim 200$~W, lasers with long beamlines, {\em e.g.} $L_1=L_2=40$~m
for ALPS \cite{Ehret:2007cm} Phase PETRA (black solid) or $L_1=L_2=170$~m for ALPS Phase
HERA (blue dotted).
Inserting ``phase-shift plates'' into the beamline as suggested in
Ref.~\cite{Jaeckel:2007gk} could improve the LSW results at larger masses.
A substantial improvement in the sensitivity to $\chi$ by several orders of
magnitude can be achieved, in the mass range from $m_{\gamma^{\prime}}\sim 10^{-7}\,\rm{eV}$ to
$m_{\gamma^{\prime}}\sim 10^{-4}\,\rm{eV}$, through experiments exploiting high-quality microwave
cavities~\cite{Jaeckel:2007ch}.
These experiments are complementary to searches for deviations of the Coulomb
law~\cite{Williams:1971ms,Bartlett:1988yy} and for photon regeneration of hidden-sector photons
produced in the Sun within the CAST magnet~\cite{Redondo} (the limit arising from the 
lifetime of the Sun is slightly worse (see also Ref.~\cite{Popov:1991})).}
\label{massive_gamma}
\end{figure}

Using the constraints on the number of photons passing through the
wall obtained from the experiments of Tab.~\ref{tab1} we can obtain new bounds in the mass-mixing plane as
shown in the left panel in Fig.~\ref{massive_gamma}.
Combining the results of the new experiments results in an improvement
by roughly a factor of two compared to the older bounds from BFRT over a wide range of masses.
In the mass range $10^{-4}\,\rm{eV} \lesssim m_{\gamma^{\prime}} \lesssim 10^{-2}\,\rm{eV}$, these
bounds are the best existing bounds on the kinetic mixing of hidden-sector photons.
(cf.~right panel in Fig.~\ref{massive_gamma}).

%%%%%%%%%%%%%%%%%%%%%%%%%%%%%%%%%%%%%
%%%%%%%%%%%%%%%%%%%%%%%%%%%%%%%%%%%%%

\section{Hidden-sector photons and minicharged particles}\label{hiddenphotons+MCP}

In Sect.~\ref{MCPs} we have simply assumed the existence of light minicharged particles without
any additional light particles being present.
However, in many models the minicharges actually arise from the coupling of a hidden-sector
particle to a hidden-sector photon that
has a kinetic mixing with the ordinary photon \cite{Holdom:1985ag}.
In other words we have minicharged particles {\emph{and}} hidden-sector photons.

Let us briefly recall how minicharged particles arise from kinetic mixing.
Assume that we have a hidden sector fermion\footnote{Here, and in the following we will
specialize
to the case where the hidden sector particle is a fermion. A generalization to scalars is straightforward
and does not change the
results qualitatively.}
$h$ that has
charge one under $B^{\mu}$. Applying the shift~\eqref{shift} to
the coupling term, we find:
\begin{equation}
e_h\bar{h}\fssd{B}\, h\rightarrow e_h\bar{h}\fssd{\tilde{B}}\, h-\chi e_h\bar{h}\fssd{A}\,
 h,
\end{equation}
where $e_h$ is the hidden sector gauge coupling.
We can read off that the hidden sector particle now has a charge
\begin{equation}
\label{epsiloncharge}
\epsilon e=-\chi e_h
\end{equation}
under the visible electromagnetic gauge field $A^{\mu}$ which has
gauge coupling $e$.
For small $\chi\ll1$, we notice that
\begin{equation}
|\epsilon|\ll 1,
\end{equation}
and $h$ becomes a minicharged particle.

However, from \Eqref{masssimple} we can see that for $m_{\gamma^{\prime}}\neq 0$ the photon propagator has
off-diagonal elements.
One finds that for momenta $q\ll m_{\gamma^{\prime}}$ these off-diagonal elements cancel the effect of the
minicharge (for details see, {\it e.g.}, Ref.~\cite{Ahlers:2007rd}) and the coupling to $h$ is effectively
zero{\footnote{To be exact, this is true only if the physical size of the magnetic field region is larger
than the inverse hidden-secton photon mass. This is always the case for the parameter regions which we are
exploring in this note. For further details see \cite{Ahlers:2007rd}. }}. In most laboratory experiments
the typical momenta are often tiny.
Therefore, if we want to have additional effects from minicharged particles the most interesting case is
that of a massless {(or nearly massless)} hidden-sector photon.

If the hidden-sector photon is massless, the photon $\leftrightarrow$ hidden-sector photon oscillations
cannot take place via a mass term as in Sect.~\ref{hiddenphotons}.
However, due to presence of the additional (light) minicharged fermions it can take place via a loop diagram
as shown in Fig.~\ref{convertBB}.
This process is now possible in addition to the production of minicharged particles as discussed in
Sect.~\ref{MCPs} that leads to an imaginary part of the
photon polarization tensor (Fig.~\ref{convertAA}).
The transition probability after a distance $z$ is \cite{Ahlers:2007rd},
\begin{equation}
P^i_{\gamma\to\gamma'}(z)=P^i_{\gamma'\to\gamma}(z)=\chi^2[1+\exp(- \kappa_{i} z/\chi^2)-
2\exp(- \kappa_{i} z/(2\chi^2)){\cos(\Delta n_{i} \omega z/\chi^2)}],
\end{equation}
where $i=\parallel,\perp$ denotes the polarization parallel or perpendicular to the magnetic field.
The total light-shining-through-a-wall probability is then,
\begin{equation}\label{transmagnetic}
P_\text{trans} = \left[
\frac{N_\text{pass}+1}{2}\right]P_{\gamma\to\gamma'}(\ell_1)P_{\gamma'\to\gamma}(\ell_2),
\end{equation}
where $\ell_1$ ($\ell_2$) denotes the length of the magnetic field in front (behind) the wall.

\begin{figure}[t]
\centering
\includegraphics[width=0.49\linewidth]{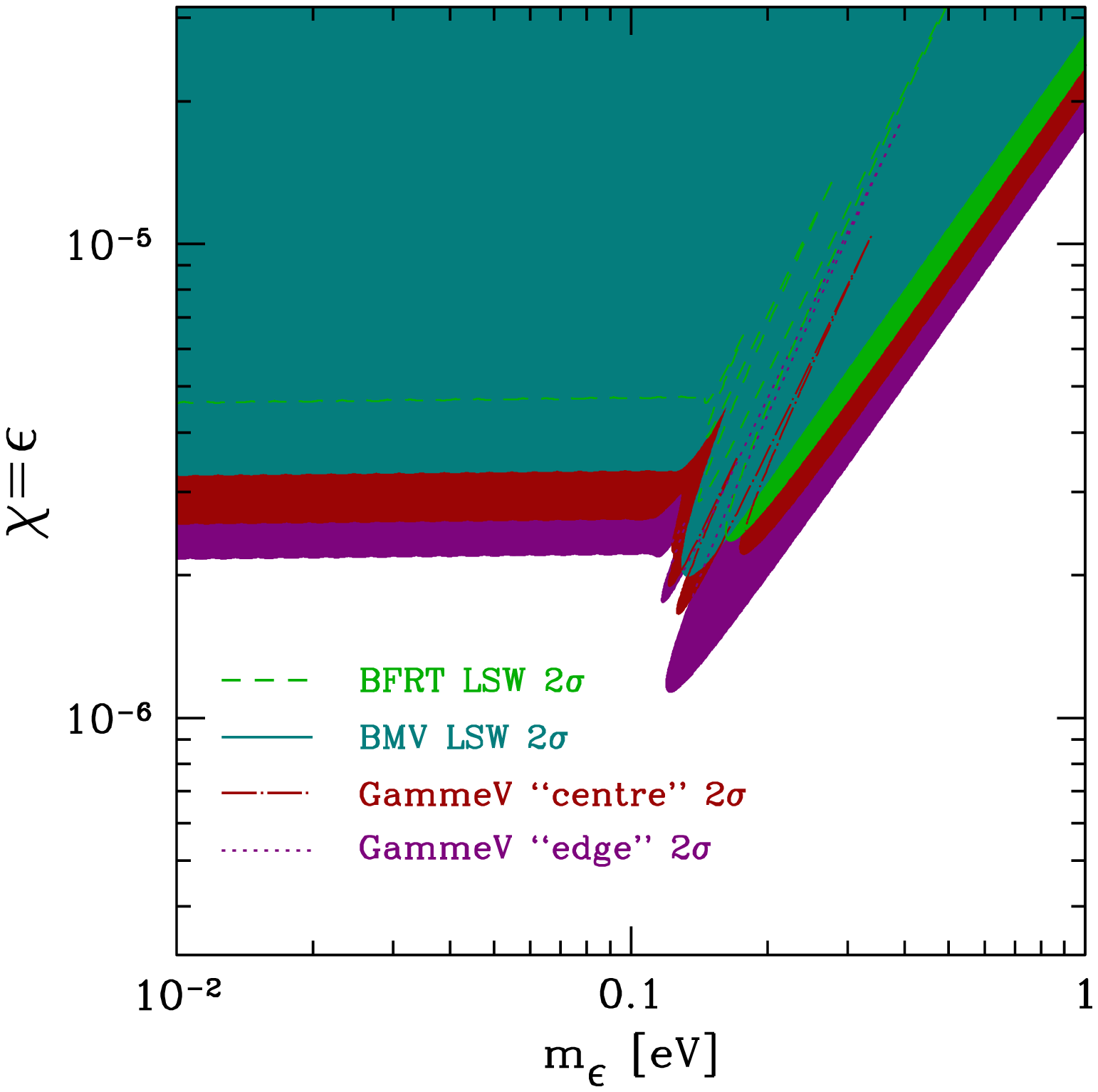}
\hfill
\includegraphics[width=0.49\linewidth]{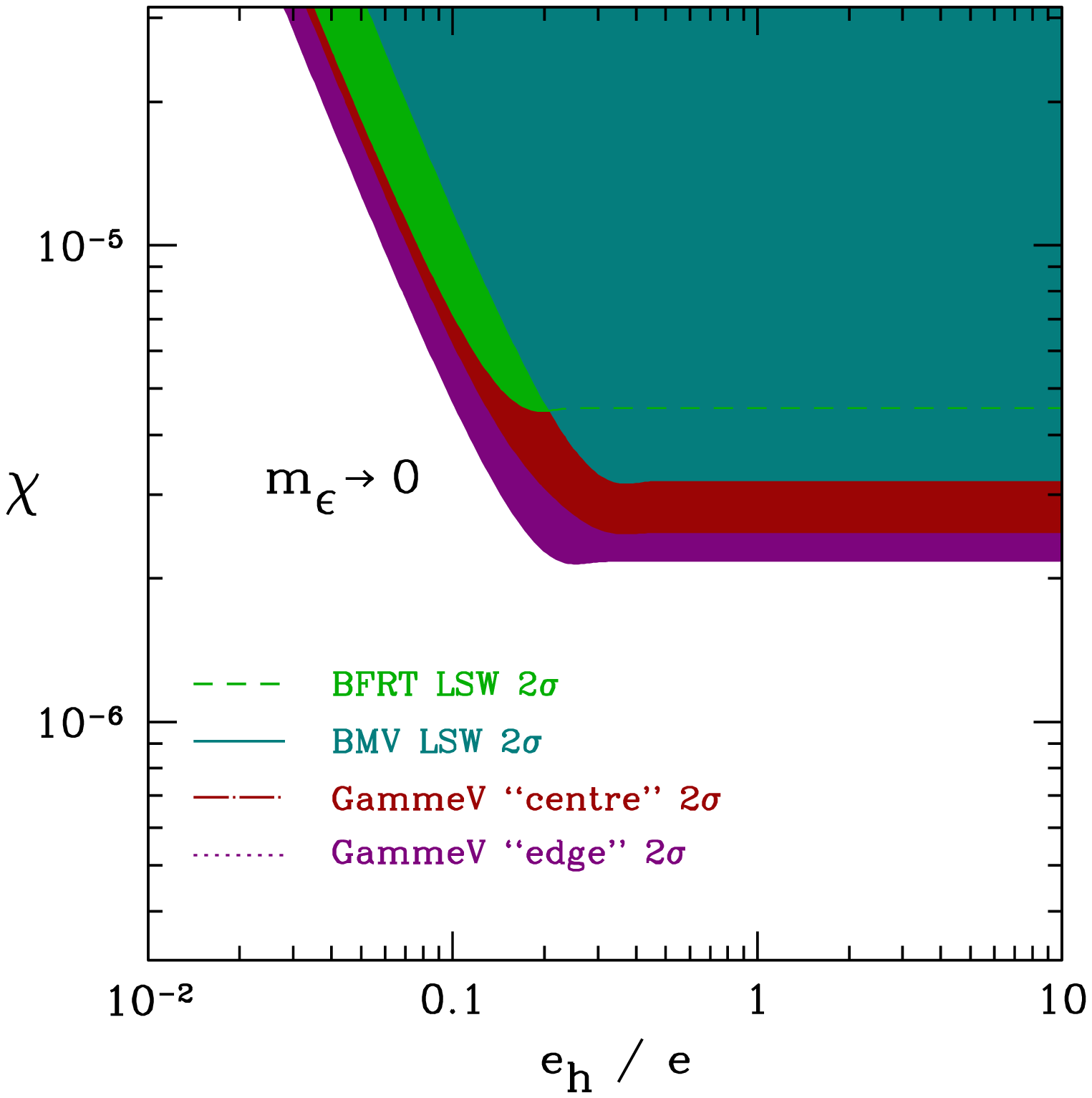}
\caption[]{\small{\bf Left panel:} Limits on the kinetic mixing of hidden-sector
photons with the ordinary electromagnetic photon in a model with a minicharged particle with
charge $\epsilon=\chi$, {\it i.e.~$e_h=-e$.} {\bf Right panel:} Limits on the kinetic mixing
parameter $\chi$ as a function
of the hidden sector gauge coupling $e_{h}$ in a model with a minicharged particle with
$m_{\epsilon}\rightarrow 0$.}\label{MCPphoton}
\end{figure}

Using the new experimental bounds, \Eqref{transmagnetic} can be turned into a bound on the possible
amount of kinetic mixing in a model with one
massless hidden-sector photon, which is shown in the left panel of Fig.~\ref{MCPphoton}
(see Tab.~\ref{tab1} for experimental parameters). The new experiments improve
the existing bounds from BFRT roughly by a factor of two. Also, for completeness we show in the
right panel
the dependence of these bounds on the {\it a priori} unknown hidden sector gauge coupling $e_{h}$ in
the particularly simple, yet enlightening, case of a massless MCP.

%%%%%%%%%%%%%%%%%%%%%%%%%%%%%%%%%%%%%
%%%%%%%%%%%%%%%%%%%%%%%%%%%%%%%%%%%%%

{Finally we would like to comment on the implications of the discussed experiments for the MR
model  \cite{Masso:2006gc}, still one of the few possibilities to evade the strong astrophysical bounds
for light MCPs. In its minimal version\footnote{In Ref.~\cite{Masso:2006gc}, the authors include also an
additional spin-zero particle.} \cite{Abel:2006qt,Ahlers:2007rd},
it features three new particles: two hidden-sector
photons, $B_\pm$, the first one mixing kinetically with the ordinary photon with mixing
parameter\footnote{The value of this mixing would be $\sqrt{2}\chi$ in \cite{Masso:2006gc}. We have
redefined it for practical purposes.} $\chi$ and the second just coupling minimally to a hidden sector
fermion $h$. If the $B_{\pm}$ photons were massless, the model would not have any phenomenological
consequences. However, adding a non-diagonal mass term $\mu^2(B_++B_-)^2/4$ in the Lagrangian
leads to
a ``link" between the photon and $h$. This link is realized as a $\mu^2$-dependent minicharge that relaxes
to zero if the mass vanishes or if it is much smaller than the other energy scales involved in the
particular process under consideration. This latter case includes the production of $h$ particles in
cosmological or astrophysical environments where the energy can be much larger than $\mu$,
for $\mu\lesssim$ eV.
In Ref.~\cite{Masso:2006gc} it is claimed that models satisfying\footnote{The authors consider 
$|e_h|=|e|$.}
$\chi\,\mu^2 \lesssim 6\times 10^{-9}\ \mathrm{eV}^2$ do not suffer from excessive anomalous energy loss
in horizontal branch stars of globular clusters, the most important constraint for the existence of sub-eV MCPs
(see however \cite{Melchiorri:2007sq}).

All these particles can show up in laser experiments as long as $\mu$ is not excessively small.
Let us focus on a LSW type of experiment\footnote{One finds (see Ref.~\cite{Ahlers:2007rd}) that rotation and ellipticity experiments are sensitive for a smaller part of the parameter space ($\chi,\mu,m_\epsilon$).}
Generically, their signature is similar to 
the signature of a hidden-photon model in such a way that we can use Fig.~\ref{massive_gamma} with $m_{\gamma^\prime}=\mu$
as a rough estimate of the exclusion bounds on the MR model as well. In fact, the MR model reduces to
the hidden photon situation developed in Sec.~\ref{hiddenphotons} for
$m_\epsilon\rightarrow\infty$. Finally,
we have checked numerically that for the opposite limiting case, namely $m_\epsilon=0$,
the combined bounds
relax by no more than a factor three in the region of Fig.~\ref{massive_gamma} (left panel).
We believe that for finite
values of $m_\epsilon$ the combined bounds should roughly lie between these two limiting cases. A detailed
study of the MR model parameter space is, however, beyond the scope of this note.}

\section{Conclusions}

Axion-like particles are the ``usual suspects'' investigated in laser polarization and
light-shining-through-a-wall (LSW) experiments. However, the non-observation of a significant signal
so far constrains also alternative scenarios, like hidden-sector photons and minicharged particles.

We have shown that the recent PVLAS limits on the birefringence and
dichroism of a magnetized vacuum constrain the charge of minicharged
particles with masses $\lesssim 0.05 \,{\rm{eV}}$ to be less than
$(3-4)\times 10^{-7}$ times the electron electric charge. This is the
best laboratory bound and comparable to bounds from the cosmic
microwave background although it is still far from the astrophysical
limits.  The latter are, however, associated with physics at a
much higher energy scale and their application to the low energy
physics probed by laboratory experiments requires an extrapolation
over many orders of magnitude.

Moreover, the LSW limits of BMV and GammeV improve the existing bounds
from BFRT on the coupling and mixing of hidden-sector photons. In the
case of massive paraphotons in the mass range $ 10^{-4}\,{\rm
eV} \lesssim m_{\gamma^{\prime}}\lesssim 10^{-2}\,{\rm eV}$, the
bound is improved by about a factor two.  These are currently the
best existing bounds on hidden-sector photons around
$m_{\gamma^{\prime}}\lesssim 10^{-3}\,\rm{eV}$ and they could be
improved even by one order of magnitude in the near future. 

Similarly, for models where minicharged particles acquire their
charge through kinetic mixing with massless hidden-sector photons, the
LSW bounds on the kinetic mixing parameter are improved by about the
same factor two.

From a general viewpoint, laser experiments have demonstrated their
capability of exploring new domains in the particle-physics parameter
space, being particularly powerful for weakly coupled light
particles. In addition to typical hidden-sector degrees of freedom,
also new fields arising in the context of cosmological models could be
searched for by similar optical techniques; see, {\it e.g.},
\cite{Ahlers:2007st,Gies:2007su}.  In the light of hidden-sector
physics, it is worthwhile to reconsider also other concepts on using
optical or more general electromagnetic signatures to deduce
information about the underlying particle-physics content, {\it e.g.},
using strong laser fields
\cite{Heinzl:2006xc,DiPiazza:2006pr,Marklund:2006my}, strong
  electric fields inside cavities \cite{Gies:2006hv}, or astronomical
observations \cite{Dupays:2005xs,Mirizzi:2007hr,DeAngelis:2007yu}.

\section*{Acknowledgements}

We thank Carlo Rizzo, Giuseppe Ruoso, and William Wester for kindly
providing us details of their light-shining-through-walls
experiments. MA acknowledges support by STFC UK (PP/D00036X/1).
HG acknowledges support by the DFG under contract No.~Gi
328/1-4 (Emmy-Noether program).

\end{document}